\documentclass[twocolumn,10 pt,showpacs,aps,prb]{revtex4-1}
\usepackage{amsmath}
\usepackage{amssymb}
\usepackage{color}
\usepackage{epsfig}
\usepackage{multirow}
\usepackage{graphicx}
\usepackage{dcolumn}

\begin{document}

\title{Surface-induced Magnetism Fluctuations in Single Crystal of NiBi$_3$
Superconductor}
\author{Xiangde Zhu,$^{1,\ast}$ Hechang Lei,$^2$ C. Petrovic,$^{2}$ and
Yuheng Zhang$^{1}$}
\affiliation{$^1$ High Magnetic Laboratory, Chinese Academy of Sciences
and University of Science and Technology of China, Hefei 230026,
People's Republic of China\\
$^2$ Condensed Matter Physics and Materials Science
Department, Brookhaven National Laboratory, Upton, New York 11973,
USA}
\date{\today}

\begin{abstract}
We report anistropy in superconducting and normal state of NiBi$_3$ single
crystals with $T_c$ = 4.06 K. The magnetoresistance results indicate the
absence of scattering usually associated with ferromagnetic metals,
suggesting the absence of bulk long range magnetic order below 300 K.
However, the electron spin resonance results demonstrate that ferromagnetism
fluctuations exist on the surface of the crystal below 150K.
\end{abstract}

\maketitle

%\begin{CJK*}{GBK}{}

\section{Introduction}

Ferromagnetism (FM) and superconductivity (SC) are two fundamental condensed
matter phenomena. The former one favors all spins parallel, while in the
latter one carriers condense in Cooper pairs where spins are antiparallel in
classical BCS scenario. The superconducting properties of ferromagnetic
superconductors are currently under debate. The $p$-wave superconductivity
was proposed to coexist with itinerant ferromagnetism. \cite{pwave} So far,
the coexistence of FM and SC has been discovered in UGe$_2$, \cite{UGe2}
URhGe, \cite{URhGe} and UCoGe. \cite{UCoGe} A binary intermetallic compound,
ZrZn$_2$ was reported to exhibit coexistence of ferromagnetism and
superconductivity. \cite{ZrZn2-nature} Interestingly, superconductivity
originates from the surface alloy rich in Zn.\cite{ZrZn2} In addition, the
spin-singlet pairing superconductivity and magnetic order parameters entwine
each other in a spatially modulated pattern, which allows for their mutual
coexistence in for example borocarbides, \cite{boroncarbides} CeCoIn$_5$ in
high magnetic fields, \cite{CeCoIn5-pwave1,CeCoIn5-pwave2} HoMo$_6$S$_8$,
\cite{HoMo6S8} P doped EuFe$_2$As$_2$, \cite{EuFe2As2} and ErRh$_4$B$_4$.
\cite{EuRh4B4}

Ni-Bi based compounds show complex physics phenomena including
antiferromagnetism and superconductivity. LaNiBiO is isostructural to $Ln$%
FeAsO ($Ln$ represents the rare earth elements) iron based high temperature
superconductors. CeNiBi$_2$ is an antiferromagnetic metal, whose structure
is similar to that of iron based superconductors. \cite{CeNiBi2}
Interestingly, when Ni deficient, CeNi$_{1-x}$Bi$_2$ shows
superconductivity. \cite{CeNiBi2} Recently, NiBi$_3$, an intermetallic
superconductor with T$_c$ = 4.06 K, attracted some attention, due to
possible coexistence of ferromagnetism (Curie temperature $<$ 750 K) and
superconductivity in polycrystals. \cite{NiBi3FM} Coexistence of
ferromagnetism and superconductivity is also observed in nano structured
crystals below 30 K. \cite{NiBi3nano} The polycrystal NiBi$_3$ is reported
to have large field dependent thermal transport properties and
magneto-resistance (MR). \cite{NiBi3MR} In addition, the magnetic ions
doping of Co results in $T_c$ enhancement and onset of resistivity drop
around 10 K in NiBi$_3$. \cite{Codoping} These results suggest that magnetic
order is related to superconductivity in NiBi$_3$. However, other two groups
reported that no magnetism was observed in NiBi$_3$ bulk crystals. \cite%
{NiBi3nano,NiBi3bulk1} It should be noted that magnetic impurities cannot be
avoided during synthesis process for nominal stoichiometric 1:3 composition
according to the Ni-Bi phase diagram. \cite{NiBi} Thus, it is necessary to
investigate the magnetism of NiBi$_3$ in impurity-free single crystals.
Here, we report the comprehensive study of NiBi$_3$. We show evidence of the
ferromagnetism fluctuations induced at the surface below 160 K. No large MR
or field dependant heat capacity is observed, which indicates that no bulk
magnetism exists in NiBi$_3$.

\section{Experimental}

Single crystals of NiBi$_3$ were grown from self-flux method with Ni$:$Bi = 1%
$:$10 mol ratio. This method avoids the NiBi phase and other magnetic
impurity in the solid state reaction with of stoichiometric Ni$:$Bi (1$:$3)
according to the Ni-Bi phase diagram. \cite{flux method} High purity Ni shot
and Bi pieces were mixed and sealed in an evacuated quartz tube. The tube
was heated to and soaked at 1150 $^\circ$C for 2 hours, then cooled down to
400 $^\circ$C with 5 $^\circ$C/h. Finally, the tube was spun in a centrifuge
to separate the Bi flux. As is shown in Fig. 1 (b), needle like, silver
colored single crystals with size of $\sim$ 2 mm $\times$ 0.2 mm $\times$
0.2 mm (with $b$ axis of the longest dimension) were obtained. The single
X-ray data were collected using the Bruker APEX2 software package Apex2 on a
Bruker SMART APEX II single crystal X-ray diffractometer with
graphite-monochromated Mo K$_\alpha$ radiation ($\lambda$ = 0.71073 \AA ) at
room temperature. The obtained lattice parameters are $a$ = 0.8955(0.0019)
nm, $b$ = 0.4153(0.0012) nm and $c$ = 1.158(0.002) nm, which is consistent
with the previous results. \cite{NiBi3bulk1} The directions of axis were
also determined.

Electrical resistance measurements were performed using a four-probe
configuration, with the applied current along the $b$ - axis. Thin Pt wires
were attached to electrical contacts made of epoxy. Electrical transport and
heat capacity measurements were carried out in Quantum Design PPMS-9. The
rotating sample holder was used to adjust the direction of the magnetic
field. During the MR measurements, $H$ is perpendicular to the $b$ - axis.
Electron spin resonance (ESR) measurement was carried out in Bruker EMX-plus
model spectrometer ($\nu \sim$ 9.4 GHz).

\section{Results and Discussions}

\begin{figure}[tbp]
\includegraphics[width=0.45\textwidth,angle=0]{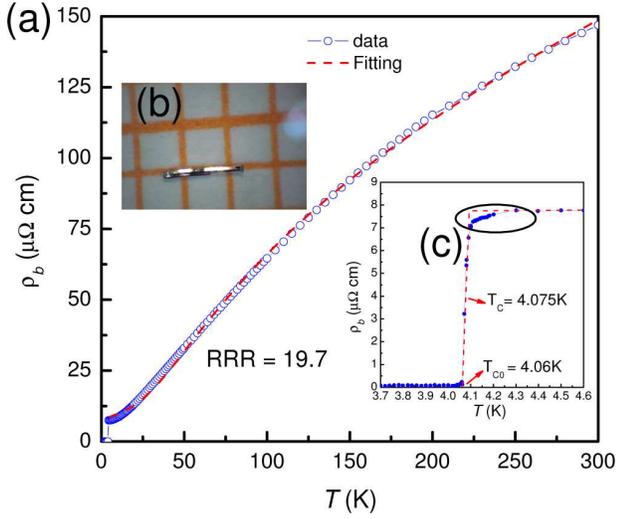}
\caption{(a) Temperature dependence of $\protect\rho_{b}$ (open circle) and
its fitted curves (dashed line) for NiBi$_{3}$. (b) The photograph of single
crystal of NiBi$_{3}$. (c) Temperature dependence of $\protect\rho_{b}$ for
NiBi$_{3}$ vicinity the superconducting transition. The ellipse marks the
initial resistivity drop. }
\end{figure}

Figure 1(a) shows the temperature ($T$) dependence of resistivity along the $%
b$-axis ($\rho_b$) for NiBi$_3$ single crystal. Its superconducting
transition can be seen in Fig. 1(c). The obtained $T_{c0}$ is about 4.06 K,
which is consistent with former results. The superconducting transition is
sharp with a transition width $\Delta T_c$ $\sim$ 0.04 K. Interestingly, the
$\rho_b - T$ curve has a initial drop around $T$ = 4.3 K (Fig. 1(c)).
Therefore, it rules out the possibility of the NiBi (S.G. $P6_3/mmc$, NiAs
type) impurity with $T_c$ = 4.25 K as the origin of the blunt specific heat (%
$C$) transition in the region from 4.1 K to 4.2 K in the polycrystalline
sample. \cite{NiBi3JPSJ} The residual resistivity ratio (RRR) is 19.7,
higher than that of the polycrystalline sample. \cite{NiBi3JPSJ} The initial
drop is reminiscent of filamentary superconductivity that precedes bulk
phase coherence, as observed in for example quasi-1D metal Nb$_2$Se$_3$.
\cite{Nb2Se3} The $\rho_b$ rises with positive curvature with increasing
temperature below 22 K, and increases with a linear $\rho_b~-~T$ relation up
to $\sim$ 60 K. Then, $\rho_b$ increases with a negative curvature above 60
K, and shows a saturation tendency towards high temperatures. The overall $%
\rho_b(T)$ follows the empirical model initially applied to A15
superconductors such as Nb$_3$Sn. \cite{WDmodel} The model gives
\begin{equation}
\rho = \rho_0+\rho_1 \cdot T+\rho_2 \cdot \exp{(\frac{-T_0}{T})}
\end{equation}
where, $\rho_0$ is the residual resistivity; $\rho_1$, $\rho_2$, and $T_0$
are material dependent parameters. The third term arises from the
phonon-assisted scattering between the two Fermi-surface sheets. The fitted
curve is shown in Fig. 1(a) as the dashed line. $\rho_0$ = 7.453$\pm$0.016 $%
\mu\Omega \cdot$ cm, $\rho_1$ = 0.244$\pm$0.004 $\mu\Omega \cdot$ cm $K^{-1}$%
, $\rho_2$ = 96.2$\pm$1.5 $\mu\Omega \cdot$ cm, and $T_0$ = 103.8$\pm$0.7 K
are determined from the fitted results. These values are comparable with
those of Nb$_3$Sn. \cite{WDmodel}

\begin{figure}[tbp]
\includegraphics[width=0.45\textwidth,angle=0]{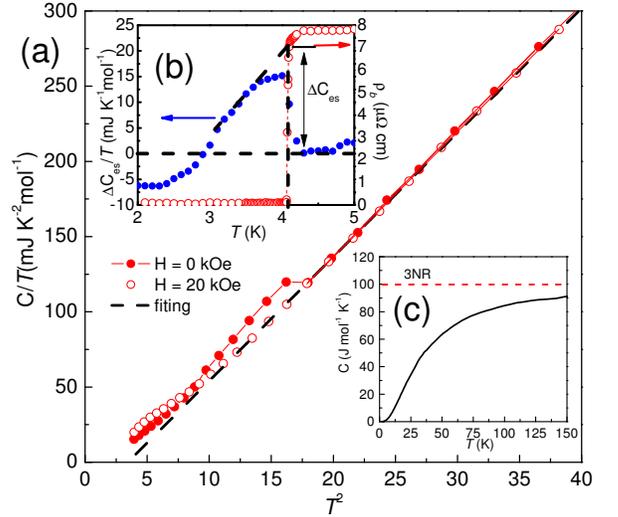}
\caption{(a) $T^2$ dependence of $C/T$ of NiBi$_3$ for $H$ = 0 kOe (solid
circle) and $H$ = 20 kOe (open circle); the linear fitting between 3 K and 6
K for $H$ = 20 kOe. (b)The $T$ dependence of $\Delta{C_{es}} = C(H = 0 kOe)
- C$($H$ = 20 kOe))(Left panel) and the $T$ dependence of $\protect\rho_b$
(right panel) for NiBi$_3$. (c)The $T$ dependence of $C$ for NiBi$_3$. The
dashed line represents the 3NR, where N and R is the number of atoms in
molecule and universal gas constant, respectively. }
\end{figure}

Figure 2(a) shows the $T^2$ dependence of specific heat ($C$) divided by ($T$%
) of NiBi$_3$ single crystal for $H$ = 0 kOe and $H$ = 20 kOe. Obviously, no
magnetic specific heat contribution can be observed. Around 4 K, a specific
heat jump due to superconducting transition can be observed. The
superconducting transition can be suppressed by $H$ = 20 kOe. Interestingly,
if we use the Sommerfeld - Debye expression $C = C_e + C_l = \gamma{T} + {%
\beta}T^3 $ ($\gamma{T}$ is the electron contribution, and ${\beta}T^3$ is
the lattice contribution) to fit the curve, the linear $C/T - T^2$ relation
between 3 K and 6 K gives unphysical negative $\gamma$. This result is
consistent with the previous reports. \cite{NiBi3JPSJ,NiBi3bulk1} Yet, $\beta
$ = 8.255$\pm$0.02 mJ mol$^{-1}$ K$^{-4}$ is obtained from the linear
fitting between 3 K and 6 K. $\Theta_D$ = 98.0$\pm$0.1K can be estimated
from the $\Theta_D = \sqrt[3]{12\pi^4{N}R/5\beta}$, where N is the number of
atoms per molecule. In addition, the $C/T - T^2$ deviates from linearity
below 3 K. Adding the anharmonic contribution $\sim T^5$ does not improve
the fit. This is most likely due to the low Debye temperature ($\Theta_{D}
\sim$ 100 K). \cite{NiBi3JPSJ} Most likely measurements of heat capacity
below 2 K are necessary in order to reliably estimate $\gamma$. At high
temperature, $C$ for NiBi$_3$ approaches the ideal value of 3NR as the
Dulong-Petit law predicts (Fig. 2(c)).

In order to investigate the specific heat jump of superconducting
transition, $C$($H$ = 0 kOe) - $C$($H$ = 20 kOe) is introduced to estimate
the electronic specific heat in the superconducting state. Figure 2(b) shows
the ${\Delta}C_{es}$($C$($H$ = 0 kOe) - $C$($H$ = 20 kOe)) and $\rho_b - T$
curves. The ${\Delta}C_{es}/T \sim$ 20 mJ mol$^{-1}$ K$^{-2}$. From the
McMillan formula
\begin{equation}
\lambda_{e-p} =\frac{\mu ^{\ast }\ln (\frac{1.45T_{c}}{\Theta _{D}})-1.04}{%
1.04+\ln (\frac{1.45T_{c}}{\Theta _{D}})(1-0.62\mu ^{\ast })},
\end{equation}
we estimate the electron-phonon coupling constant $\lambda_{e-p}$ $\sim$
0.91 by assuming $\mu ^{\ast }$ = 0.13, which is a typical value for the
Coulomb pseudo-potential. \cite{mu} This indicates that NiBi$_3$ is a
strongly electron-phonon coupled superconductor.

Semiclassical transport theory predicts that Kohler's rule will be valid if
there is a single species of charge carrier and the scattering time $\tau$
is the same at all points on the Fermi surface. MR at different temperatures
can be scaled by the expression eq.(3) with the assumption that scattering
rate $1/\tau$ is proportional to $\rho(T)$, and $\omega_c$ is the frequency
at which the $H$ causes the charge carriers to sweep across the Fermi
surface:
\begin{equation}
\frac{\Delta\rho_{xx}(H,T)}{\rho_{xx}(0,T)} = f{(\omega_c\tau)} = f{(\frac{H%
}{\rho(0,T)})},
\end{equation}
and the corresponding plots are known as Kohler's plots.

\begin{figure}[tbp]
\includegraphics[width=0.45\textwidth,angle=0]{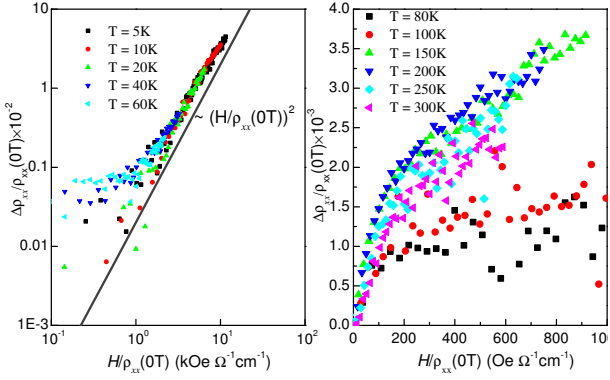}
\caption{(a) MR the Kohler's log-log plot in single crystals of for NiBi$_{3}
$ at $T$ = 5, 10, 20, 40, 60 K. (b) MR at different temperatures and the
Kohler's plot in single crystals of for NiBi$_{3}$ at $T$ = 80, 100, 150,
200, 250, 300 K. }
\end{figure}

Figure 3(a) depicts the Koler's plot for NiBi$_3$ from 5 K to 60 K. Even at $%
H \perp b$ = 9 T, the MR is only about 4.5$\%$ at $T$ = 5 K. This value is
very typical for a nonmagnetic metal. All curves collapse onto a single
line, suggesting that the Kohler's rule is valid ($\Delta\rho_{xx}(H)$/$%
\rho_{xx}$(0) $\propto (\mu_0H/\rho_{xx}$(0))$^2$). Figure 3(b) depicts the
MR data $\Delta\rho_{xx}(H)$/$\rho_{xx}$(0) as a function of $%
\mu_0H/\rho_{xx}$(0) above 80 K. Apparently, Kohler's rule is violated above
80 K. Due to the crystal structure, NiBi$_3$ should be a quasi-one
dimensional system, whose Fermi surface topology is not spherical. The
possible explanation is that multiband conductivity in NiBi$_3$. Below 60 K,
the contribution of one band (with large $\omega_c\tau$) dominates whereas
the contribution of other bands is negligible. The system can then be
regarded as a single band conductor. Above 80 K the contributions are
comparable, leading to deviation of the Kohler's rule. Similar phenomenon
has been observed in low dimensional system, such as LiFeP. \cite{LiFeP}
What is important, the MR is positive below 300 K, indicating that NiBi$_3$
is not ferromagnetic in bulk.

\begin{figure}[tbp]
\includegraphics[width=0.45\textwidth,angle=0]{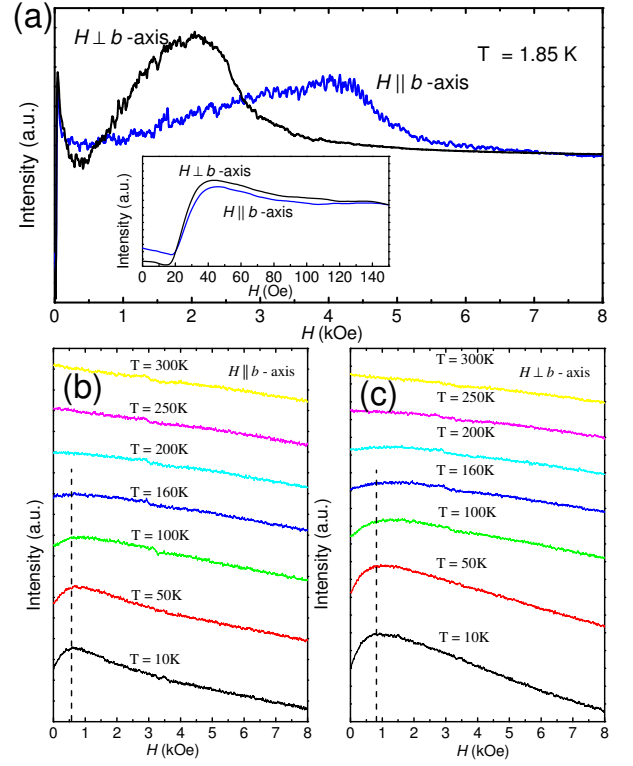}
\caption{(a)The ESR spectra of NiBi$_3$ measured at 1.85 K with sweeping $H$
parallel to the $b$ - axis and perpendicular to $b$-axis. Inset shows the
amplified plots in the low field region. The ESR spectra of NiBi$_3$
measured at different temperatures with sweeping $H$ parallel to the $b$ -
axis (b) and perpendicular to $b$ - axis (c). The curves are shifted for
clarity. The dashed lines mark the maximum of the ESR spectrums. }
\end{figure}

Figure 4(a) shows the ESR ($dP/dH$) spectra of NiBi$_3$ measured at 1.85 K
with sweeping $H$ parallel to the $b$ - axis ($H \parallel b$) and
perpendicular to $b$-axis ($H \perp b$). Obviously, anisotropy can be
observed. Both spectra show typical ESR signals for supercondcuting state.
As shown in the inset, a stepwise sharp signal appears at low fields typical
for the magnetic-shielding feature, indicating the appearance of
superconductivity. Then, ESR signal show a hump below $H_{c2}$ with
increasing $H$.

Figure 4 (b) and (c) show the ESR spectra of NiBi$_3$ measured at different
temperatures (above $T_c$) with sweeping $H \parallel b$ and $H \perp b$,
respectively. No obvious resonance signal around $H$ = 3200 Oe for
paramagnetic electron can be observed. Interestingly, the ESR spectra of NiBi%
$_3$ show abnormal diplike signal at low field below 160 K, which can be
suppressed by increasing temperature. As marked by the dashed lines in Fig.
4 (b) and (c), the magnetic fields corresponding to diplike signal are
anisotropic. In heavy fermion system CeRuPO, FM correlations is define by
the ESR signal. \cite{ESR-FM} At low temperature below 10 K (Curie
temperature = 15 K), ESR of CeRuPO shows similar diplike signals at low
fields as well as NiBi$_3$. This diplike signal should be related to
ferromagnetic moment fluctuations. At high temperature, the diplike signal
of NiBi$_3$ disappears. The microwave signal of ESR can only exist on the
surface of a metal (in $\mu$m length). \cite{microwave} This strongly
suggests that ferromagnetic moments fluctuations exist the surface on NiBi$_3
$ below 160 K, and that NiBi$_3$ is non-magnetic in bulk. The crystal
structure of NiBi$_3$ can be regarded as packing of NiBi$_3$ rods along the $%
b$ - axis. In the bulk, the coordination of NiBi$_3$ rods has perfect
translational symmetry; while on the surface, the translational symmetry is
broken. The FM fluctuations in NiBi$_3$ could be attributed to the surface
effect since surface tension modifies the electronic band structure,
favoring FM on submicron length scale in nanostructured samples. \cite%
{NiBi3nano}

\begin{figure}[!htb]
\includegraphics[width=0.45\textwidth,angle=0]{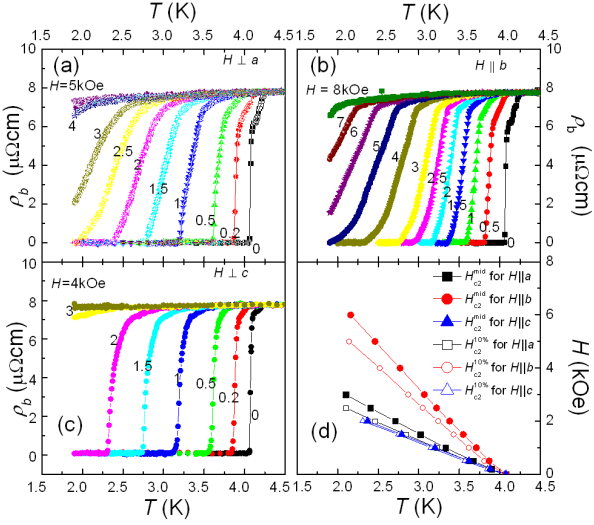}
\caption{Temperature dependence of the resistivity $\protect\rho_b$($T$) of
NiBi$_3$ for (a) $H \parallel a$, (b) $H \parallel b$ and (c) $H \parallel c$
a at the various magnetic fields. (d) Temperature dependence of the upper
critical field $H_{c2}$ obtained from the midpoint and 10\% on the $\protect%
\rho_b - T$ curves for $H \parallel a$, $H \parallel b$ and $H \parallel c$.
}
\end{figure}

\begin{table}[!htb]
\caption{Parameters for NiBi$_3$.}
\label{TableKey}\centering%
\begin{tabular}{cccccc}
\hline\hline
$T_{c0}$ &  & \multicolumn{4}{c}{4.06 K} \\
$\rho_0$ &  & \multicolumn{4}{c}{7.453$\pm$0.016 $\mu\Omega \cdot$ cm} \\
$\rho_1$ &  & \multicolumn{4}{c}{0.244$\pm$0.004 $\mu\Omega \cdot$ cm $K^{-1}
$} \\
$\rho_2$ &  & \multicolumn{4}{c}{96.2$\pm$1.5 $\mu\Omega \cdot$ cm} \\
$T_0$ &  & \multicolumn{4}{c}{103.8$\pm$0.7 K} \\
$\rho_{sat}$ &  & \multicolumn{4}{c}{328$\pm$2 $\mu\Omega$ cm} \\
$A_{ac}$ &  & \multicolumn{4}{c}{0.886$\pm$0.001 $\mu\Omega \cdot$ cm $K^{-1}
$} \\
$\lambda_{e-p}$ &  & \multicolumn{4}{c}{$\sim$ 0.91} \\
$\Theta_D$ &  & \multicolumn{4}{c}{98.0$\pm$0.1K} \\
$H_c(0)$ &  & \multicolumn{4}{c}{$\sim$ 430 Oe} \\ \cline{3-5}
&  & $a$ & $b$ & $c$ &  \\ \cline{3-5}
\multirow{2}*{$d{H_{c2}}/d{T|_{T_c}}$} & mid & 1.48 & 3.13 & 1.17 & kOe/K \\
& 10\% & 1.22 & 2.60 & 1.14 & kOe/K \\
\multirow{2}*{$H_{c2}$(0)} & mid & 6.0 & 12.7 & 4.75 & kOe \\
& 10\% & 4.95 & 10.56 & 4.62 & kOe \\
\multirow{2}*{$H_{c1}$(0)} & mid & 65 & 41 & 75 & Oe \\
& 10\% & 71 & 47 & 74 & Oe \\
\multirow{2}*{$\xi$(0)} & mid & 18.1 & 38.3 & 14.3 & nm \\
& 10\% & 18.3 & 39.0 & 17.1 & nm \\
\multirow{2}*{$\kappa$(0)} & mid & 9.9 & 20.9 & 7.8 &  \\
& 10\% & 8.14 & 17.4 & 7.6 &  \\
\multirow{2}*{$\lambda_{GL}$(0)} & mid & 180 & 800 & 112 & nm \\
& 10\% & 149 & 678 & 130 & nm \\
\multirow{2}*{$m_a:m_b:m_c$} & mid & \multicolumn{3}{c}{4.48 : 1 : 7.15} &
\\
& 10\% & \multicolumn{3}{c}{4.58 : 1 : 5.02} &  \\ \hline\hline
\end{tabular}%
\end{table}

Figure 5 (a), (b), and (c) show the $\rho_b - T$ curves at the various
magnetic fields for $H \parallel a$, $H \parallel b$ and $H \parallel c$,
respectively. The upper critical field ($H_{c2}$) is obtained from the
midpoint and 10\% resistivity on the $\rho_b - T$ curves, which is depicted
in Fig. 5 (d). Obviously, $H_{c2}$ show linear temperature dependence in the
experimental temperature range. The obtained $[dH_{c2}^{mid}/dT]|_{T_c}$ ($%
[dH_{c2}^{10\%}/dT]|_{T_c}$) are 1.48(1.22), 3.13(2.60), and 1.17(1.14)
kOe/K for $H \parallel a$, $H \parallel b$, and $H \parallel c$,
respectively. According to the conventional one-band
Werthamer-Helfand-Hohenberg (WHH) theory, \cite{WHH} which describes the
orbital limited $H_{c2}$ of dirty type-II superconductors, $H_{c2} =
0.693[dH_{c2}/dT]|_{T_c}T_c$. Then $H_{c2}^a$(0), $H_{c2}^b$(0), $H_{c2}^c$%
(0) can be obtained as 6(4.95), 12.7(10.56), and 4.75(4.62) kOe. The data in
the brackets here and the following discussion are within the 10$\%$
resistivity results.

In the anisotropic superconducting system, coherence length ($\xi_{GL}^i$,
where $i$ is the axis) can be estimated from $H_{c2}^a = \frac{\Phi_0}{%
2\pi\xi_b\xi_c}, H_{c2}^b = \frac{\Phi_0}{2\pi\xi_a\xi_c}, H_{c2}^c = \frac{%
\Phi_0}{2\pi\xi_a\xi_b}$. The calculated $\xi_{GL}^a$, $\xi_{GL}^b$, $%
\xi_{GL}^c$ are 18.1(18.3) nm, 38.3(39.0) nm, and 14.3(17.1) nm,
respectively. The effective mass tensor is related to the GL coherence
length as $m_a:m_b:m_c$ = $1/{\xi_{GL}^a}^2:1/{\xi_{GL}^b}^2:1/{\xi_{GL}^c}^2
$. The calculated effective mass ratio $m_a:m_b:m_c$
4.48:1:7.15(4.58:1:5.02). Obviously, NiBi$_3$ shows a quasi one dimensional
behavior along the $b$ - axis, with effective mass ratio $m^*_a \simeq m^*_c$
and $m^*_i/m^*_b \simeq$ 5 ($i$ represents $a$ and $c$).

The thermodynamic critical magnetic field at $T$ = 0 K ($H_c(0)$) can be
estimated through $H_c(0)=[4\pi N(E_F)\Delta^2(0)]^{1/2}$. According to the
band calculation, the density of states at Fermi surface $N(E_F)$ is 2.55
states/eV f.u. \cite{NiBi3bulk1} By assuming the $\Delta(0) = 1.76 k_BT_c$, $%
H_c$ is estimated to be $\sim$ 430 Oe. This value is consistent with
experimental results. \cite{NiBi3bulk1} GL parameter $\kappa$, the lower
critical field at zero temperature ($H_{c1}$(0)), and the penetration depth $%
\lambda_{GL}$ at $T$ = 0 K can be estimated from: $H_{c2}^i = \sqrt{2}%
\kappa^i{H_c}^i, H_{c1}^i = H_c\frac{\ln{\kappa^i}-0.18}{\sqrt{2}\kappa^i},
and~~\lambda_{GL}^i = \kappa^i{\xi_{GL}^i}$, where $i$ represents the
direction of $a, b, c$. The estimated parameters in the superconducting
state and the parameters in the normal state mentioned above are listed in
Tab. I.

\section{Summary}

We investigated anisotropy in superconducting and normal state properties of
NiBi$_3$. The carrier scattering mechanism is dominated by electron-phonon
scattering whereas MR suggests the absence of magnetic scattering commonly
observed in bulk FM materials. The heat capacity results indicates it is a
strongly $e-p$ coupling superconductor. However, FM fluctuations are
detected at the surface of the crystal. Finally, we give anisotropic
superconducting parameters.

\section{Acknowledgement}

We thank Dr. Wei Tong for the help on ESR measurements. Work at Brookhaven
National Laboratory (H. L. and C. P.) was supported by the US DOE under
Contract No. DE-AC02-98CH10886. Work at High magnetic field lab (Hefei) was
supported by the State Key Project of Fundamental Research, China
(2010CB923403) and National Basic Research Program of China (973 Program),
No. 2011CBA00111.

* Email:xdzhu@hmfl.ac.cn

%\end{CJK*}


\begin{thebibliography}{99}
\bibitem{microwave} Skin effect depth = $\sqrt{\rho/\pi f \mu}$, where $\rho$
is the resistivity of the conductor in $m$, $f$ is frequency in Hertz,
and $\mu$ is the absolute magnetic permeability of the conductor. The
absolute magnetic permeability $\mu$ = $\mu_0 \cdot \mu_r$, where $\mu =
4\pi \times 10^{-7}$ H/m, and $\mu_r$ is the relative magnetic permeability.
Since NiBi$_3$ is not magnetic in bluk, here we use $\mu_r$ = 1 to estimate
the Skin effect depth. As for NiBi$_3$, $f$ = 9.4Ghz, $\rho$ = 8 - 150 $\times
10^{-8}$ ohm m. The skin depth should be 1.5 - 6.4 $\mu$m.
\end{thebibliography}
\end{document}